# The Inherent Behavior of Graphene Flakes in Water: A Molecular Dynamics Study

Priyanka Solanky[1], Vidushi Sharma[2,*], Kamalika Ghatak[2,*], Jatin Kashyap[2], Dibakar Datta[2]

[1] 12th grade student, Montville Township High School, Montville, New Jersey
[2] Department of Mechanical and Industrial Engineering, New Jersey Institute of Technology (NJIT), Newark, New Jersey

[*] Corresponding Authors

Vidushi Sharma, Email – vs574@njit.edu, Phone - 862 872 8687
Kamalika Ghatak, Email – kg0602@njit.edu, Phone – 734 548 2812

## *Abstract*

Graphene- water interaction has been under scrutiny ever since graphene discovery and realization of its exceptional properties. Several computational and experimental reports exist that have tried to look into the interactions involved, however, none of them addresses the issue in its entirety. Most computational analysis doesn't go beyond the atomistic scale and report graphene to be hydrophobic. Meanwhile, several experimental and molecular dynamics (MD) studies show an active interaction between graphene and water. We have, therefore, tested the inherent hydrophobic behavior of a small graphene in water droplet by the means of MD simulations. The analysis has been extended to multiple graphene flakes in water and their respective size dependent responses to water droplet. Graphene retreats from water droplet to encapsulate it from the surface. This response was highly dependent upon graphene size with respect to water content. Additionally, we also report self-assembly of multilayered graphene in water by means of MD simulations, an observation which can be utilized to synthesize such structures in a cost-effective way by experimentalists. To fully comprehend graphene behavior in water, graphene deformation was analyzed in the presence of water molecules. It was noticed that graphene wrinkled to wrap around water molecules and resisted complete failure, one that is seen in case of a sole graphene sheet. Our work will not only address the question about whether graphene is hydrophobic or hydrophilic but also provide insight into the behavior of graphene surface and mobility when exposed to water which can be exploited in numerous applications.

## I. INTRODUCTION

In the realm of two dimensional(2D) materials, single atom thick Graphene made of $sp^2$ hybridized carbons arranged in honeycomb lattice has dominated research ever since its rediscovery in 2004.[1] This is primarily due to the groundbreaking properties that make it a very promising material in all sphere of science and technology. Graphene is gifted with exceptional properties in terms of electronics[1], optics[2], thermal conductivity[3], mechanical strength[4], and very high surface area. Above all else, it is mostly surface properties that render graphene potent in widespread applications in the field of biomedical sciences[5], micro-fluidics[6-7], bio-imaging[8], batteries[9] to name a few. Surface properties of graphene are highly controlled by its morphology, chemical additives, and synthesis techniques. While first isolation of graphene was done using scotch tape, multiple techniques for graphene synthesis have been devised in past two decades such as chemical exfoliation[10], chemical vapor deposition (CVD)[11], suction filtration[12] and bottom-up synthesis (Tang-Lau Method[13]). CVD is prominent method to synthesize graphene in vitro and it is a common knowledge that substrate used in CVD affects overall response to water of synthesized graphene surface.[14] Interaction of graphene with water has been a long-standing question in the research community due to the ambiguity of its character.

Based on first principle calculations by Leenaerts et. al.[15], graphene was initially concluded to be hydrophobic. From density functional theory(DFT) results, they established that a single water molecule in a cluster does not bind strongly to graphene sheet (adsorption energy per molecule of water 13 meV) and charge transfer between the water molecule and graphene sheet decreases as water cluster size increases. Following this report, many applications of graphene were explored in nanoelectromechanical and microelectromechanical systems where a possibility of liquid deposition was needed to be reduced[16], in microfluidics, catalysis, electrochemistry, corrosion and gas sensing. [5-7, 17] On the other hand, it was around the same time as Leenaerts et. al.[15] published their results, there was another DFT report which showed water molecules adsorbed on graphene surface and desorption was not feasible at room temperature.[18] It was realized that such contradictory graphene-water interaction is due to complex combinations of electrostatic, H-bonding, and van der Waals interactions. Further experimental and molecular dynamics reports were published discussing reversibility of graphene's hydrophobicity to hydrophilicity by several mechanisms including substrate controlled synthesis, altering ultraviolet radiation [19], chemical additives[20]. Needless to say, graphene water interaction can be engineered based on the requirement of the application. All the previous studies based on DFT, molecular dynamics and experiments detail closely upon atomic interactions between oxygen and hydrogen of water with the carbons of graphene and their effect on electronic, structural and chemical properties of latter. However, an intrinsic behavior of a graphene when exposed to the large quantity of water or humidity has not been emphasized enough beyond the atomistic scale. In this regard, it has been previously reported that graphene placed in a water nano-droplet is pushed to the surface driven by its hydrophobicity and forms van der Waals interactions with water molecules to encapsulate them.[21] The major reason for such surface film adsorption is given to be a reduction in surface free energy on water adsorption. This Molecular Dynamics (MD) report has been backed up by experimental observations.

Therefore, in this work, the idea has been extended to multiple graphene sheets in water and their respective interactions. The hydrophobic attributes of graphene flakes($G_f$) are investigated by means of MD simulations. Single and multiple $G_f$ are placed in the water droplet, and the inherent behavior of graphene is examined. To be precise, relativity between a size of graphene and water content is investigated. Graphene-water interaction is largely directed by the size of a graphene sheet and content of water. Further, for the first-time humidity directed self-assembly of multilayered graphene (MLG) has been shown. These assembled graphene sheets were aligned at different turbostratic orientations, and their hydrophobicity was amplified with each addition of graphene layers. Additionally, deformation behavior of graphene in water has been predicted based on its stress-strain curve.

## II. METHODOLOGY

### *2.1 Molecular Structures*

A cubic simulation box of side 150 Å with hemispheric water droplet having diameter 120 Å and the number of water molecules $N_w \sim 21000\text{-}30000$ were constructed for this work. In order to model graphene-water size dependent interactions, three cases were considered: **(i)** Three representative small $G_f$ of size ~45 Å were placed in the water droplet at offset positions such that two of three $G_f$ were slightly exposed outside of the water droplet (see figure 1(a)) while the central one remains completely in water. **(ii)** Another system was considered where all three $G_f$ were exposed outside the water droplet (see figure 1(b)). The difference between each $G_f$ in the z-direction was taken to be 40 Å. **(iii)** Three large $G_f$ with the surface area comparable to that of water content were taken as shown in figure 1(c). The dimension of large square graphene are around ~90Å and distance between two $G_f$ in z direction is taken to be 33Å. To simulate retreat of single $G_f$ from water, a single $G_f$ of dimensions ~42Å was constructed such that one-third of the flake is already exposed outside spherical water droplet (figure 1(d)). This was done in order to speed the retreat process of $G_f$ from the water. Further to model graphene-graphene interaction in presence of water content several $G_f$ (seven) were placed in water at offset position such that they were slightly exposed (see figure 1(e)) with side ~45Å and difference between two flakes were kept 13Å in z direction. To extend the analysis, about 20 $G_f$ were placed in water droplet at offset position with size ~ 20Å and 4Å distance in z direction (see figure 1(f)). To study the stress-strain relation of graphene in presence of water, a periodic graphene sheet was taken of dimensions about 126X121Å in a simulation box(126X121X100 Å) full of water molecules. Two cases were considered: Tensile loading was applied to **(i)** a periodic graphene, and **(ii)** a cracked or previously deformed graphene in a box full of water. The slit-like crack with armchair edges was taken of length $2a$ ($a = 10$Å) in the middle of graphene. Dimensions of graphene and slit length were chosen in order to avoid effects of finiteness.

### *2.2 Molecular Dynamics simulations*

In order to model carbon-carbon bonds, Adaptive Intermolecular Reactive Bond Order (AIREBO) potential[22] was used in Large-scale Atomic/Molecular Massively Parallel Simulator (LAMMPS) package[23] for the MD simulations of graphene in water where interaction cut-off parameter was taken as 1.92 Å. TIP4P water model[24] was used to model all the other bonds in the system. Periodic boundary conditions were maintained for the system. All the structures were minimized via conjugate gradient (cg) method in LAMMPS pre-MD run. Simulations were run in NVE ensemble with Berendsen thermostat to maintain the temperature at room temperature (300 K) with a damping constant of 100 fs. Time-step for each simulation run was taken to be of 2 fs. The final simulation time varies from 3-6 ns depending upon the system.

Tensile loading was applied to graphene sheet in x direction at a fixed strain rate of 0.001 until complete failure at room temperature under NPT ensemble. Stress on the graphene sheet was calculated by averaging the stress on all the carbon atoms over the graphene sheet. Stress on individual carbon atoms was calculated by Virial theorem[25] considering graphene to be of thickness 3.4 Å:

$$\sigma_{ij}^{\alpha} = \frac{1}{\Omega}\left(\frac{1}{2}m^{\alpha}v_i^{\alpha}v_j^{\alpha} + \sum_{\beta=1,n}\gamma^j f^i\right)$$

where i,j denote indices in Cartesian system; α,β are atomic indices; $m^{\alpha}$ and $v^{\alpha}$ denote mass and velocity of atom α; γ is the distance between atoms α and β; $f$ is the force applied by atom α on β and Ω is the volume of an atom α. Stress- strain curve was determined as done in the previous study by Pei et.al.[26]

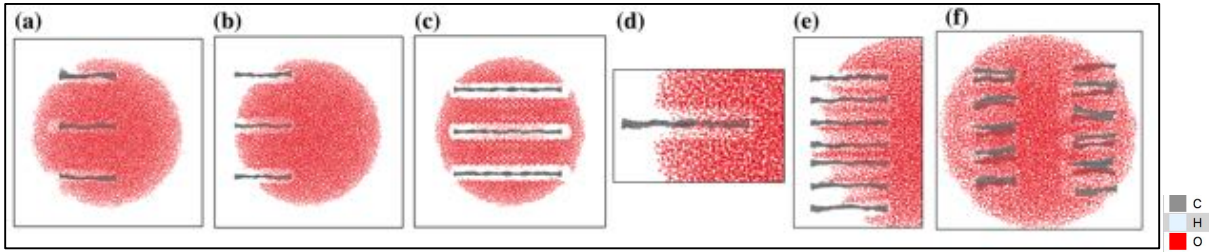

*Figure 1*: Cross-sectional view of starting structures of graphene flake($G_f$) in a spherical water droplet for molecular dynamics (MD) simulations at 0 timestep.

## III. RESULTS AND DISCUSSION

*3.1. Graphene-water size-dependent interaction*

Density Functional Theory studies to predict the adsorption of water on graphene have shown ambiguous results.[15, 18, 27-29] Due to π electrons of graphene, it is highly sensitive to environment especially to the humidity.[28] Water is known to withdraw electrons from graphene, therefore having a direct influence on its electronic properties.[30] This makes Graphene-water interaction a crucial factor that influences the performance of graphene in many potential applications. Graphene monolayer is well known to be hydrophilic[18]; however, its behavior varies between hydrophobic to hydrophilic based on its method of synthesis, stacking[18, 31-32], substrate[27] etc. As previously mentioned, graphene has been reported to drive its way out of the water nano-droplet and form van der Waals interactions with water molecules to encapsulate them.[21] The idea has been extended to multiple graphene sheets in water to study their respective behavior in presence of other graphene sheets. Here, several monolayer $G_f$ were kept inside the water droplet and simulated for an extended amount of time using MD to investigate the inherent tendency of $G_f$ in the presence of water and other $G_f$. $G_f$ were placed in a water droplet at different positions and simulations were run for 6 ns. Three cases were considered as already discussed in section 2.1 with varying $G_f$ size and position. It was noted that $G_f$ and water interaction was greatly dependent upon the size of $G_f$.

For cases where small $G_f$ were placed at an offset position with respect to water droplet, two scenarios were considered: one where two of three $G_f$ are exposed outside water content and another where all three were slightly exposed outside. It was seen that hydrophobic tendency of $G_f$ dominated as they retreated from water during the run. In figure 2(a-b) it can be seen that $G_f$ are placed at slightly offset position in water initially. The top and bottom graphene came out quickly and covered the water surface. The tendency of graphene to retreat from water increases when exposed outside the water droplet. This suggests hydrophobic character dominates hydrophilic character for $G_f$. On the other hand, the middle graphene with no initial exposure continued to stay in the water. No difference in the system was noticed post this point for extended MD run. This can be due to domineering encapsulating and gliding $G_f$ outside which prevent surfacing of central graphene from the water. In case **(ii)** (figure 2(c-d)), $G_f$ were placed slightly more offset such that all three graphene are exposed to the air. All three of them manage to retreat from the water. While the top and bottom graphene form an arc covering the water droplet, central graphene initiated formation of double layered graphene by interacting with the nearest graphene present. This phenomenon also assisted in fast retreating of graphene from the water. This implies that the tendency to form a stacked structure dominates hydrophilicity of graphene and hydrophobic nature increases with stacking as previously reported.[31-32] This is more detailed in section 3.3.

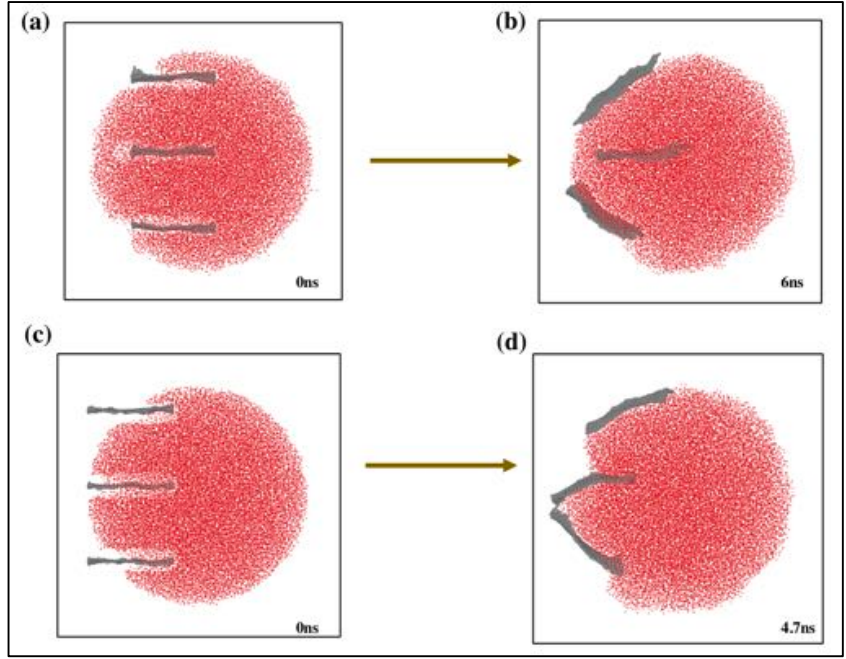

*Figure 2*: Cross-sectional view of small graphene flakes ($G_f$) in a water droplet before and after simulation. Position of $G_f$ in water influence the process of their out-coming and overlapping. **(a-b)** $G_f$ with respect to water after 0 ns and 6 ns of MD simulation for case **(i)**. The simulation was run for longer time but no difference in system was noticed post this point. **(c-d)** $G_f$ with respect to water after 0 ns and 4.7 ns of MD simulation when all three $G_f$ were slightly exposed outside water in case **(ii).**

For comparatively larger graphene (case **(iii)**), as shown in figure 3(a), though inherently hydrophobic, $G_f$ did not retreat from water but rather formed an encapsulating curvature around prominent part of the water. This entrapment of water molecules in between graphene sheets is in accordance with the results published by Song et. al. where they exhibited that there is a prominent charge transfer taking place between water molecules and graphene sheets when they are placed in between two graphene sheets.[33] It can be seen in figure 3(c) that top and bottom graphene form an arc around water while central graphene does not move appreciably. Rest of the water molecules form a droplet interacting with least amount of graphene surface area as possible. Another interesting observation during the simulation run was the orientation re-arrangement of $G_f$ with respect to one another. At the beginning of the simulation (see Figure 3 (d); top view) , $G_f$ were placed exactly on the top of one another, however, during and after the run, they were shifted from the initial position and were found to align themselves at slight angles with respect to one another (see Figure 3(e); top view). It is well-known that multilayered graphene (MLG) is more stable at AB stacking order[34] which explains the re-arrangement of $G_f$ with respect to one another.

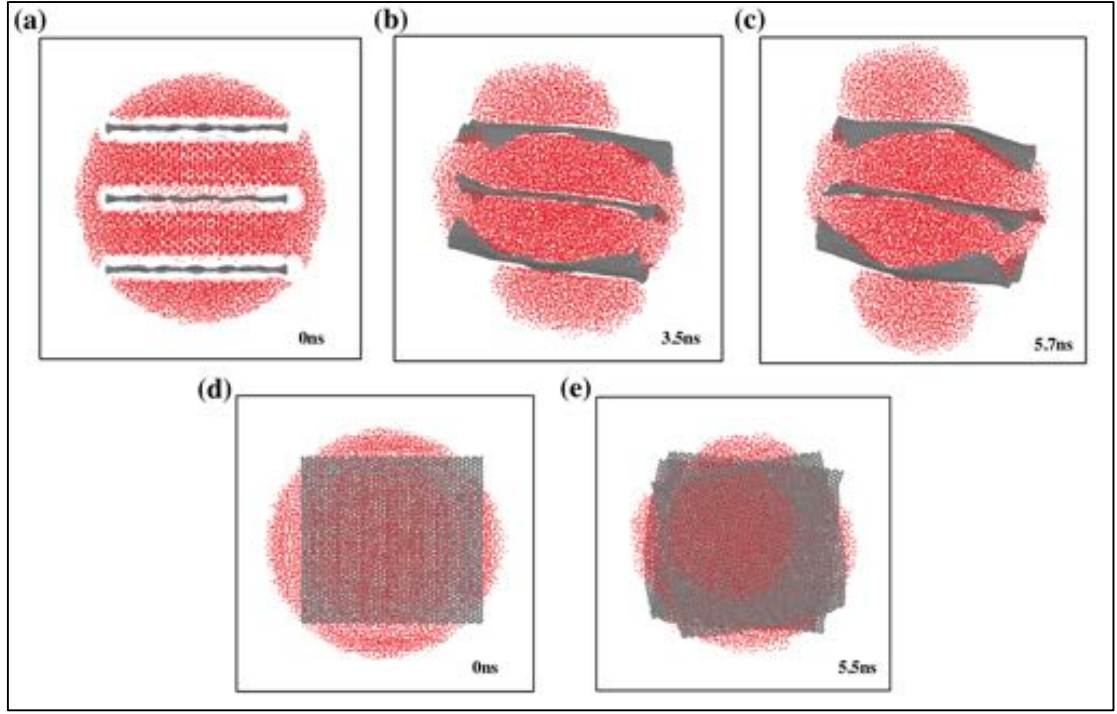

*Figure 3*: **(a-c)** Cross-sectional view of graphene flakes($G_f$) in a water droplet of comparable cross-sectional area during simulation at different timesteps from start to end (0 - 5.7 ns). **(d-e)** Top view of starting and final graphene-water system. Orientation re-arrangement of graphene flakes can be seen. The top view remained consistent from ~3 ns throughout the simulation run.

## *3.2. Retreat of graphene from water at the surface*

In the case of multiple graphene in water, we observed that graphene flakes interacted with each other to form a stacked structure as shown in figure 2(d). In order to map the retreat of single graphene flake from water in absence of other controlling parameters, we stimulated a single graphene flake in the water droplet. $G_f$ was placed in the water droplet such that one-third of the flake is already exposed outside (figure 1(d)). This was done in order to speed the retreat process of $G_f$ from water as in the previous studies graphene has been reported to take significant time to come out of the water.[21] Additionally, as we have already seen in section 3.1 case **(i)** $G_f$ with no initial exposure may not retreat from water in presence of other $G_f$ that are encapsulating the water drop. Nevertheless, in all the graphene-water simulations done in this study, it was seen that $G_f$ eventually comes out of the water and encapsulates the water droplet on the surface. It can be seen in figure 4 which shows a cross-sectional view of $G_f$ coming out of the water droplet during the simulation run and covering the water content from outside. To the best of our knowledge, the process of graphene retreat from water surface has not been discussed before. Initially, the surface of the out-coming $G_f$ is perpendicular to the water droplet surface. The $G_f$ faces strong water adsorption at the base and therefore bends from the intersection towards water surface to form van der Waals interactions. These week forces induce gliding of graphene at the surface. In the process, the contact angle between $G_f$-water reduces, and the flake continues to come out of water simultaneously covering the drop surface. This process was observed during simulations of multiple $G_f$ as discussed in section 3.1.

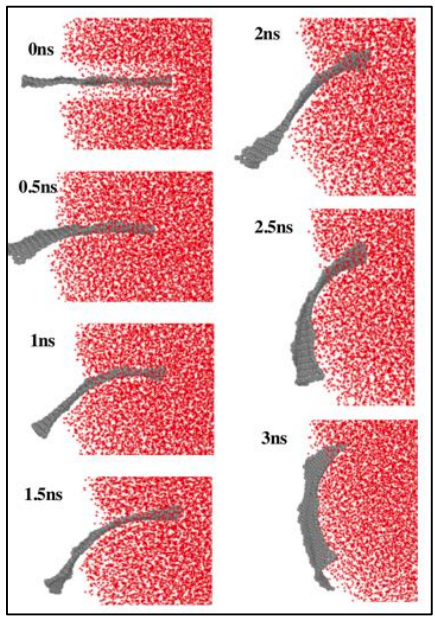

*Figure 4:* Cross-sectional view of retreat of a graphene flake ($G_f$) from water during MD simulation taken after every 0.5ns of simulation run.

*3.3. Self-assembly of Multilayered Graphene*

The response of graphene to water strongly depends upon its size and position with respect to water content as seen in section 3.1 and 3.2. However, during the simulation of three $G_f$ in water, we observed that graphene during its retreat from water interacts with nearby graphene to form a stacked structure (figure 2(d)). In order to extend our analysis to include the effect of several $G_f$ on the system, a system with seven $G_f$ and twenty $G_f$ were taken and simulations were run. It was noticed that graphene flakes assembled to form multilayered graphene. Multilayer Graphene (MLG) is a stacked form of Graphene up to 5-10 graphene layer that lack three-dimensional order which is usually seen in case of Graphite.[35] In simple words, atomic positions of carbon in one plane have no relation with the ones in other planes. However, if the layer number exceeds 10, such materials usually exhibit properties similar to Graphite. In light of two most dominant carbon structures, i.e., graphene and graphite, MLG lie less explored in terms of their properties and applications. High surface area and mechanical strength of MLG makes it demanding material for several graphene based applications. While synthesis of MLG via conventional methods is expensive, it was observed from our study that graphene flakes have the tendency to self-assemble into multilayered structures in water. As shown in figure 5, seven $G_f$ were placed in water and simulated for ~6ns. The process of $G_f$ retreat from the water was accompanied by their simultaneous stacking. This process of stacking further accelerated the progression of graphene coming out of water increasing the hydrophobic character of MLG. The number of graphene sheets in the formed MLG varied from 2 to 5. It is to be pointed out that no external force or chemical additive was added into the system during simulations. To the best of our knowledge, no previous report of MLG formation driven by simply water hydrophobicity exists. From our observation, it is clear that the interaction between two graphene sheets dominated the interaction between water and graphene. Further, we simulated twenty $G_f$ in water. The starting structure with twenty $G_f$ within water droplet is shown in figure 6(a) and figure 6(b) shows the multilayered stacked structures which eventually either retreat from water and cover the water droplet surface or drift away from water. It was seen that MLG up to 3 layers of graphene covered the surface of water droplet after each layer had completely come out of the water. While MLG with more than 3 layers of graphene drifted away from water exhibiting strong hydrophobicity.

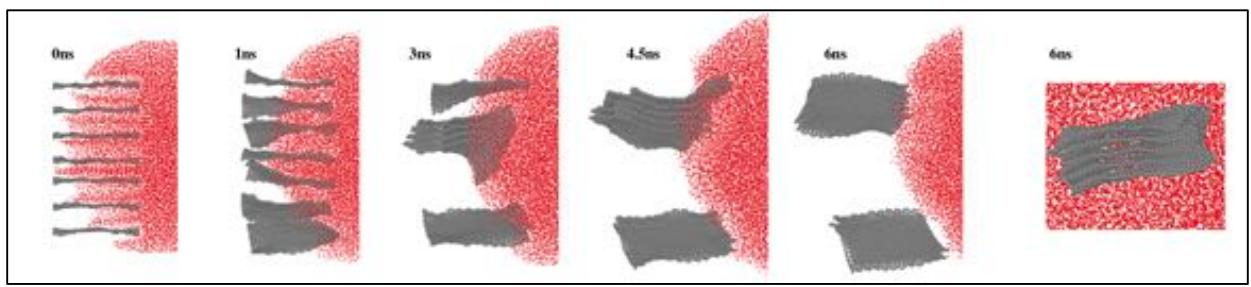

*Figure 5:* Cross-sectional view of retreat of seven graphene flakes ($G_f$) from water during simulation to form multilayered graphene (**MLG**) by stacking together. Images are taken at different timestep during simulation run.

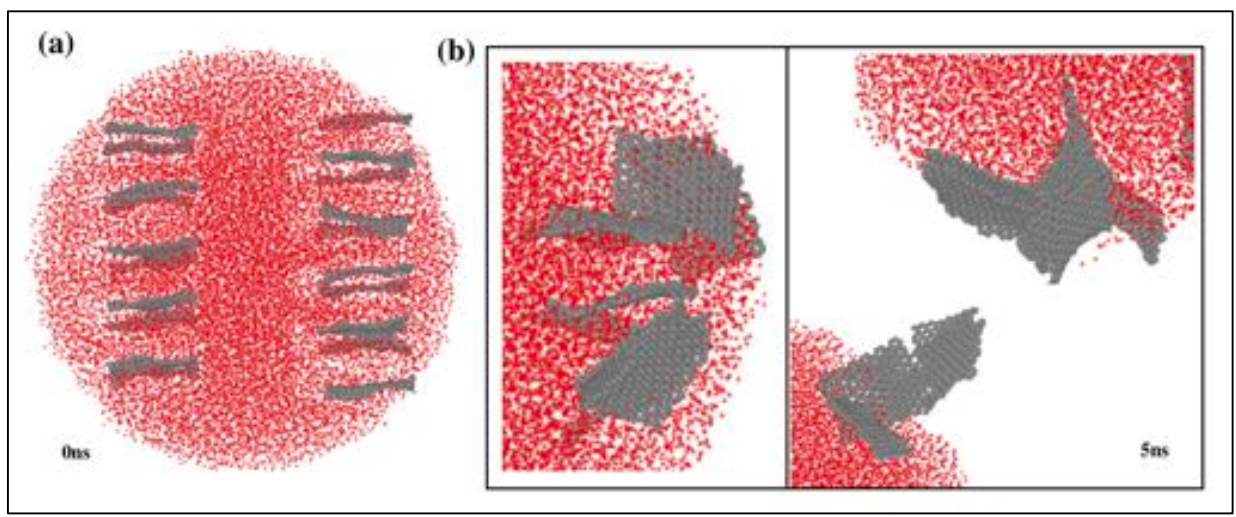

*Figure 6:* (*a*) Cross-sectional view of twenty graphene flakes ($G_f$) placed in water before simulation. (*b*) Stacking of $G_f$ during simulation run to form stacked multilayered structures. Images are taken after 5ns of simulation run.

*3.4 Effect of water on stress-strain response of graphene*

Stability of single graphene sheet in abundant water content calls for some additional analysis as this trait can be well exploited in terms of its biomedical and sensing applications. Graphene currently is the known strongest material which can withstand stresses of about ~100-130 GPa without failure.[26] Once the crack is initiated in graphene, it undergoes brittle failure. In the light of current standing in terms of graphene-water interaction, we can say that graphene exhibits interestingly mixed traits where its hydrophilicity and hydrophobicity are greatly dependent upon its size and water content. In order to predict deformation of graphene in water, the stress-strain response of graphene in a simulation box full of water was done in LAMMPS by applying a constant strain rate of 0.001. Two cases were considered as previously mentioned : Tensile loading was applied to **(i)** a periodic graphene, and **(ii)** a cracked or previously deformed graphene in a box full of water. This was done in order to apprehend effect of water molecules on defected graphene sheet and its subsequent performance under stress. The comparison was made with graphene in the simulation box with no water under similar stress keeping the dimensions consistent.

For the case **(i)** when tensile loading was applied to pristine graphene in a simulation box full of water molecules (see figure 7(a)), it was seen that during the deformation, water molecules in the entire simulation box accumulated in the central region and graphene sheet wrinkled to wrap around the water molecules as can be seen in figure 7(b). This observation was in accordance with the analysis done in section 3.1 where large graphene flakes encapsulated water content. Figure 8 represents the stress-strain plot for graphene in presence of water in comparison with solely graphene. It can be seen in figure 8, even though graphene fractures at a strain of 0.17, when placed in water, it does not undergo complete fragmentation but only a small crack at strain of 0.177. The plausible reason for such a behavior can be attributed to the severe wrinkling of graphene which has been reported previously to increase its fracture toughness.[36] As a result, while there was a crack nucleation at a strain of 0.177, no subsequent brittle fracture was seen. Additionally, no further straining was possible for the system. We hypothesize that water-graphene interactions and wrinkled topology have a huge role to play in resisting further straining. The investigation regarding which is still ongoing.

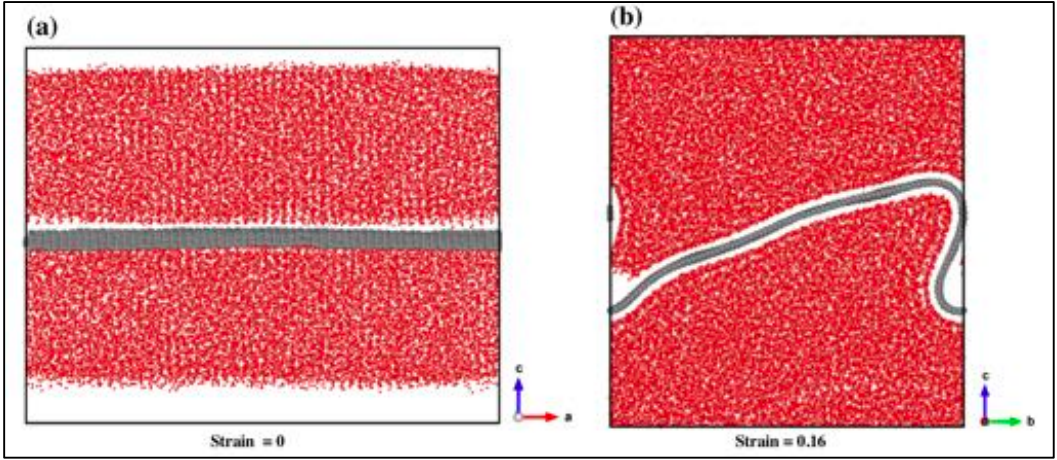

*Figure 7:* **(a)** Side view of a simulation box full of water molecules and a periodic graphene sheet present at the center at the onset of tensile loading. **(b)** Horizontal view of simulation box from x direction under tensile loading. Wrinkling of graphene sheet around the water molecule can be seen.

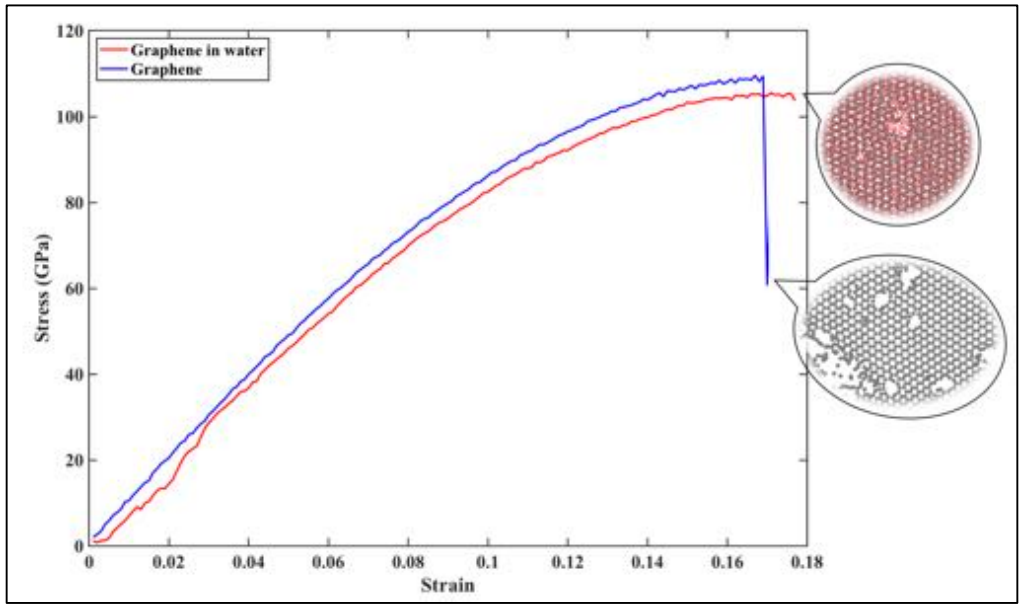

*Figure 8:* The stress- strain plots of graphene in water (red) and graphene without water (blue). Small crack in graphene sheet at highest strain (0.177) for which simulation ran has been shown contrary to completely fractured graphene sheet at strain 0.17 in absence of water molecules.

In the second case **(ii)**, tensile loading was applied to a graphene in a box full of water having a slit-like crack with armchair edges of length $2a$ ($a = 10$Å) in the middle. Similar to the case **(i)**, during deformation, water molecules accumulated in center of the box and graphene sheet wrinkled to wrap around them. Figure 9(a) shows stress-strain plot for cracked graphene with and without water and response of graphene to water on deformation was very similar to one seen in case of pristine graphene. While the atomic stresses were slightly smaller in presence of water, the maximum strain that could be applied was very close (0.118 for graphene in water and 0.122 for the only graphene). Again while graphene fractured completely at strain 0.122, in presence of water no further straining beyond 0.118 strain was possible due to excessive wrinkling which resisted deformation. Our main focus was to notice the process of crack propagation if any in presence of water. It can be seen in figure 9(b-d), the slit-like crack was in the process of healing during the process and no crack propagation was seen. Slight re-arrangement of bonds near around the slit was seen throughout.

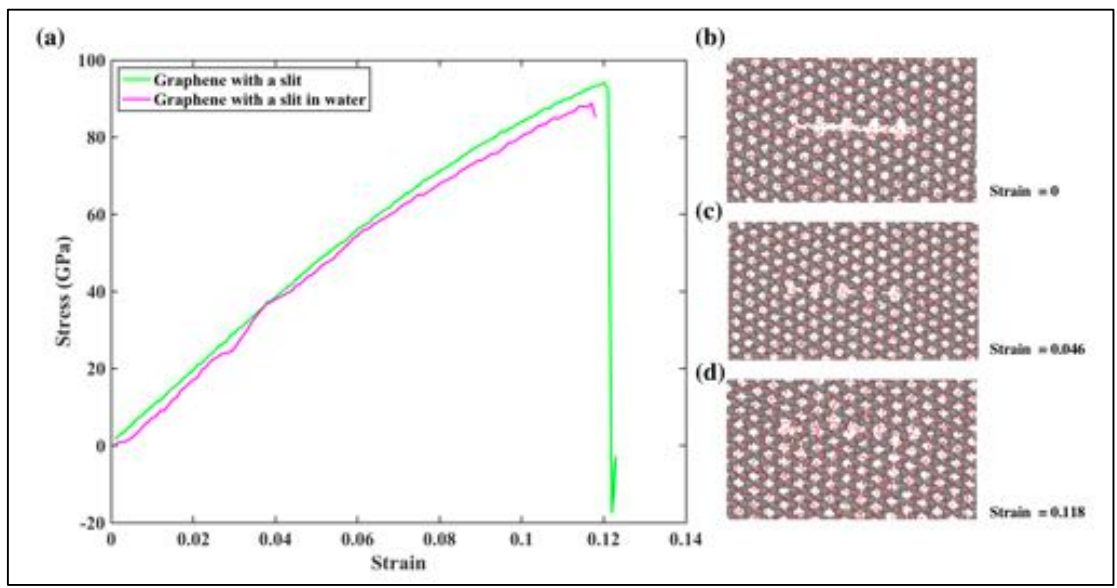

*Figure 9:* **(a)** The stress- strain plots of cracked graphene in water (magenta) and cracked graphene without water (green). **(b-d)** Top view of slit in graphene during deformation, depicting re-arrangement of bonds.

## IV. CONCLUSION

To conclude, an attempt has been made in this study to understand the response of free-standing graphene when placed in water. With extensive use of graphene in applications involving water-based solutions, especially in biomedical sciences, it is necessary to predict surface trait of such 2D material when exposed to water content. We performed molecular dynamics simulations of different sized and number of graphene in water in order to apprehend the size and number dependent response of graphene flakes to water. We found that graphene tends to cover the surface area of the water content. Large graphene encapsulates the water, while smaller graphene flakes first retreat from the water and then form surface interaction with water content. It is known that graphene can be hydrophobic or hydrophilic depending on numerous environmental conditions. We found supporting evidence to this theory during our simulations. When graphene was exposed to the air, hydrophobicity dominates that drive graphene outward from the water content. While once it is out, it adsorbs water molecules on its surface and covers the water surface. In addition to this, for the first time, we showed by means of molecular dynamics simulations that water can drive several graphene sheets to self-assemble to form a multilayered graphene. These multilayered graphene sheets were found to be hydrophobic as they do not adsorb water molecules on the surface, an attribute which increases with the number of graphene layers. It was noted that graphene-graphene interactions dominated over graphene-water interaction. Our study highlights mixed traits of graphene with respect to water content which can be well utilized by the research community in terms of its applications, especially in the field of drug delivery as graphene can be used to encapsulate drug emulsions. Further, we hypothesize based on our simulations that MLG can be produced in a cost-effective way by self-assembling in water. These self-assembled MLG were hydrophobic and were stacked at different orientations with respect to one another. In addition, a small attempt has been made to address the deformation behavior of graphene in water. It was found that during tensile loading, graphene sheet wrinkles to wrap around water molecule and does not undergo brittle failure. The strain at which graphene is supposed to fracture, we see nucleation of small cracks. This is mostly because of graphene wrinkling and its interaction with water. Further investigations are currently being undertaken in this regard.

**Acknowledgement**
VS and KG acknowledge the financial support from the NJIT. We are grateful to the High-Performance Computing (HPC) facilities managed by Academic and Research Computing Systems (ARCS) in the Department of Information Services and Technology (IST) of the New Jersey Institute of Technology (NJIT). Some computations were performed on Kong HPC cluster, managed by ARCS. We acknowledge the support of the Extreme Science and Engineering Discovery Environment (XSEDE) for providing us their computational facilities (Start Up Allocation - DMR170065 & Research Allocation - DMR180013). Most of these calculations were performed in XSEDE SDSC COMET Cluster.

**Data Availability**

The raw/processed data required to reproduce these findings cannot be shared at this time as the data also forms part of an ongoing study. However, it will be shared by the authors upon request.

V. REFERENCES

1. Novoselov, K. S.; Geim, A. K.; Morozov, S. V.; Jiang, D.; Zhang, Y.; Dubonos, S. V.; Grigorieva, I. V.; Firsov, A. A., Electric field effect in atomically thin carbon films. *science* **2004,** *306* (5696), 666-669.
2. Kuzmenko, A.; Van Heumen, E.; Carbone, F.; Van Der Marel, D., Universal optical conductance of graphite. *Physical review letters* **2008,** *100* (11), 117401.
3. Balandin, A. A.; Ghosh, S.; Bao, W.; Calizo, I.; Teweldebrhan, D.; Miao, F.; Lau, C. N., Superior thermal conductivity of single-layer graphene. *Nano letters* **2008,** *8* (3), 902-907.
4. Lee, C.; Wei, X.; Kysar, J. W.; Hone, J., Measurement of the elastic properties and intrinsic strength of monolayer graphene. *science* **2008,** *321* (5887), 385-388.

5. Shen, H.; Zhang, L.; Liu, M.; Zhang, Z., Biomedical applications of graphene. *Theranostics* **2012,** *2* (3), 283.

6. Sui, S.; Wang, Y.; Kolewe, K. W.; Srajer, V.; Henning, R.; Schiffman, J. D.; Dimitrakopoulos, C.; Perry, S. L., Graphene-based microfluidics for serial crystallography. *Lab on a Chip* **2016,** *16* (16), 3082-3096.

7. Ang, P. K.; Li, A.; Jaiswal, M.; Wang, Y.; Hou, H. W.; Thong, J. T.; Lim, C. T.; Loh, K. P., Flow sensing of single cell by graphene transistor in a microfluidic channel. *Nano letters* **2011,** *11* (12), 5240-5246.

8. Zhu, S.; Zhang, J.; Qiao, C.; Tang, S.; Li, Y.; Yuan, W.; Li, B.; Tian, L.; Liu, F.; Hu, R., Strongly green-photoluminescent graphene quantum dots for bioimaging applications. *Chemical communications* **2011,** *47* (24), 6858-6860.

9. Kim, H.; Park, K.-Y.; Hong, J.; Kang, K., All-graphene-battery: bridging the gap between supercapacitors and lithium ion batteries. *Scientific reports* **2014,** *4*, 5278.

10. Stankovich, S.; Dikin, D. A.; Piner, R. D.; Kohlhaas, K. A.; Kleinhammes, A.; Jia, Y.; Wu, Y.; Nguyen, S. T.; Ruoff, R. S., Synthesis of graphene-based nanosheets via chemical reduction of exfoliated graphite oxide. *carbon* **2007,** *45* (7), 1558-1565.

11. Reina, A.; Jia, X.; Ho, J.; Nezich, D.; Son, H.; Bulovic, V.; Dresselhaus, M. S.; Kong, J., Large area, few-layer graphene films on arbitrary substrates by chemical vapor deposition. *Nano letters* **2008,** *9* (1), 30-35.

12. Zhang, X.; Wan, S.; Pu, J.; Wang, L.; Liu, X., Highly hydrophobic and adhesive performance of graphene films. *Journal of Materials Chemistry* **2011,** *21* (33), 12251-12258.

13. Tang, L.; Li, X.; Ji, R.; Teng, K. S.; Tai, G.; Ye, J.; Wei, C.; Lau, S. P., Bottom-up synthesis of large-scale graphene oxide nanosheets. *Journal of Materials Chemistry* **2012,** *22* (12), 5676-5683.

14. Kim, G. T.; Gim, S. J.; Cho, S. M.; Koratkar, N.; Oh, I. K., Wetting-Transparent Graphene Films for Hydrophobic Water-Harvesting Surfaces. *Advanced Materials* **2014,** *26* (30), 5166-5172.

15. Leenaerts, O.; Partoens, B.; Peeters, F., Water on graphene: Hydrophobicity and dipole moment using density functional theory. *Physical Review B* **2009,** *79* (23), 235440.

16. Zong, Z.; Chen, C.-L.; Dokmeci, M. R.; Wan, K.-t., Direct measurement of graphene adhesion on silicon surface by intercalation of nanoparticles. AIP: 2010.

17. Yuan, W.; Shi, G., Graphene-based gas sensors. *Journal of Materials Chemistry A* **2013,** *1* (35), 10078-10091.

18. Ribeiro, R.; Peres, N.; Coutinho, J.; Briddon, P., Inducing energy gaps in monolayer and bilayer graphene: local density approximation calculations. *Physical Review B* **2008,** *78* (7), 075442.

19. Yang, J.; Zhang, Z.; Men, X.; Xu, X.; Zhu, X., Reversible superhydrophobicity to superhydrophilicity switching of a carbon nanotube film via alternation of UV irradiation and dark storage. *Langmuir* **2010,** *26* (12), 10198-10202.

20. Rafiee, J.; Rafiee, M. A.; Yu, Z. Z.; Koratkar, N., Superhydrophobic to superhydrophilic wetting control in graphene films. *Advanced Materials* **2010,** *22* (19), 2151-2154.
21. Chen, Y.; Guo, F.; Jachak, A.; Kim, S.-P.; Datta, D.; Liu, J.; Kulaots, I.; Vaslet, C.; Jang, H. D.; Huang, J., Aerosol synthesis of cargo-filled graphene nanosacks. *Nano letters* **2012,** *12* (4), 1996-2002.
22. Stuart, S. J.; Tutein, A. B.; Harrison, J. A., A reactive potential for hydrocarbons with intermolecular interactions. *The Journal of chemical physics* **2000,** *112* (14), 6472-6486.
23. Plimpton, S., Fast parallel algorithms for short-range molecular dynamics. *Journal of computational physics* **1995,** *117* (1), 1-19.
24. Jorgensen, W. L.; Chandrasekhar, J.; Madura, J. D.; Impey, R. W.; Klein, M. L., Comparison of simple potential functions for simulating liquid water. *The Journal of chemical physics* **1983,** *79* (2), 926-935.
25. Basinski, Z.; Duesbery, M.; Taylor, R., Influence of shear stress on screw dislocations in a model sodium lattice. *Canadian Journal of Physics* **1971,** *49* (16), 2160-2180.
26. Pei, Q.; Zhang, Y.; Shenoy, V., A molecular dynamics study of the mechanical properties of hydrogen functionalized graphene. *Carbon* **2010,** *48* (3), 898-904.
27. Wehling, T. O.; Lichtenstein, A. I.; Katsnelson, M. I., First-principles studies of water adsorption on graphene: The role of the substrate. *Applied Physics Letters* **2008,** *93* (20), 202110.
28. Leenaerts, O.; Partoens, B.; Peeters, F., Adsorption of H 2 O, N H 3, CO, N O 2, and NO on graphene: A first-principles study. *Physical Review B* **2008,** *77* (12), 125416.
29. Freitas, R.; Rivelino, R.; Mota, F. d. B.; De Castilho, C., DFT studies of the interactions of a graphene layer with small water aggregates. *The Journal of Physical Chemistry A* **2011,** *115* (44), 12348-12356.
30. Surwade, S. P.; Smirnov, S. N.; Vlassiouk, I. V.; Unocic, R. R.; Veith, G. M.; Dai, S.; Mahurin, S. M., Water desalination using nanoporous single-layer graphene. *Nature nanotechnology* **2015,** *10* (5), 459-464.
31. Kozbial, A.; Li, Z.; Sun, J.; Gong, X.; Zhou, F.; Wang, Y.; Xu, H.; Liu, H.; Li, L., Understanding the intrinsic water wettability of graphite. *Carbon* **2014,** *74*, 218-225.
32. Munz, M.; Giusca, C. E.; Myers-Ward, R. L.; Gaskill, D. K.; Kazakova, O., Thickness-dependent hydrophobicity of epitaxial graphene. *Acs Nano* **2015,** *9* (8), 8401-8411.
33. Song, R.; Feng, W.; Jimenez-Cruz, C. A.; Wang, B.; Jiang, W.; Wang, Z.; Zhou, R., Water film inside graphene nanosheets: electron transfer reversal between water and graphene via tight nano-confinement. *RSC Advances* **2015,** *5* (1), 274-280.
34. Shibuta, Y.; Elliott, J. A., Interaction between two graphene sheets with a turbostratic orientational relationship. *Chemical Physics Letters* **2011,** *512* (4-6), 146-150.
35. Bianco, A.; Cheng, H.-M.; Enoki, T.; Gogotsi, Y.; Hurt, R. H.; Koratkar, N.; Kyotani, T.; Monthioux, M.; Park, C. R.; Tascon, J. M., All in the graphene family–a recommended nomenclature for two-dimensional carbon materials. Elsevier: 2013.
36. Qin, H.; Sun, Y.; Liu, J. Z.; Liu, Y., Mechanical properties of wrinkled graphene generated by topological defects. *Carbon* **2016,** *108*, 204-214.